# Domain evolution in bended freestanding BaTiO$_3$ ultrathin films: a phase-field simulation

Changqing Guo,[1] Guohua Dong,[2] Ziyao Zhou,[2] Ming Liu,[2] Houbing Huang,[3,4,a)] Jiawang Hong,[1,a)] and Xueyun Wang[1,a)]

[1]School of Aerospace Engineering, Beijing Institute of Technology, Beijing 100081, China

[2]Electronic Materials Research Laboratory, Key Laboratory of the Ministry of Education, School of Electronic and Information Engineering, and State Key Laboratory for Mechanical Behavior of Materials, Xi'an Jiaotong University, Xi'an 710049, China

[3]School of Materials Science and Engineering, Beijing Institute of Technology, Beijing 100081, China

[4]Advanced Research Institute of Multidisciplinary Science, Beijing Institute of Technology, Beijing 100081, China

[a)] Electronic mail: xueyun@bit.edu.cn; hongjw@bit.edu.cn; hbhuang@bit.edu.cn.

**Abstract**

Perovskite ferroelectric oxides are usually considered to be brittle materials, however, recent work [Dong *et al.*, Science 366, 475 (2019)] demonstrated the super-elasticity in the freestanding BaTiO$_3$ thin films. This property may originate from the ferroelectric domain evolution during the bending, which is difficult to observe in experiments. Therefore, understanding the relation among the bending deformation, thickness of the films, and the domain dynamics is critical for their potential applications in flexible ferroelectric devices. Here, we reported the dynamics of ferroelectric polarization in the freestanding BaTiO$_3$ ultrathin films in the presence of large bending deformation up to 40° using phase-field simulation. The ferroelectric domain evolution reveals the transition from the flux-closure to *a*/*c* domains with "vortex-like" structures, which caused by the increase of out-of-plane ferroelectric polarization. Additionally, by varying the film thickness in the identical bending situation, we found the *a*/*c* phase with "vortex-like" structure emerges only as the film thickness reached 12 nm or higher. Results from our investigations provide instructive information for the microstructure evolution of bending ferroelectric perovskite oxide films, which could serve as guide for the future application of ferroelectric films on flexible electronic devices.

Ferroelectric materials with reversible spontaneous electrical polarization and the corresponding ferroelectric domain structures[1-4], are ideal mediums for nonvolatile memories, thin-film capacitors, and actuators, *etc*[5-10]. With the development of modern intelligent devices, achieving better flexibility is the new trend of ferroelectric materials[11-14] to meet the demand for flexible and wearable devices. Up to date, ferroelectric oxides with perovskite structure have been extensively studied on their physical, chemical, electrical, and magnetic properties[15-18]. In general, perovskite oxides are considered as brittle and non-bendable materials, which greatly limits the applications of ferroelectric oxides in flexible electronics and smart wearable devices.

Recently, a few seminal works demonstrated that super-elasticity was also achievable in such ferroelectric perovskite. For example, single crystalline $BaTiO_3$ (BTO) freestanding films[19] and $Pb(Mg_{1/3}Nb_{2/3})O_3$-$PbTiO_3$[20] can undergo extremely large bending, especially for the freestanding BTO film, a 180° bending can be realized without any hint of cracks. The striking phenomenon provides an additional degree of freedom for the application of ferroelectric materials. Though the authors suggested that micro-ferroelectric domains are responsible for the super-elasticity, ferroelectric domain evolution, as well as the corresponding change of ferroelectric properties in the presence of large bending deformation remain unknown due to the challenges in the experiments. The phase-field method can be used to simulate the evolution of ferroelectric thin films under external loading or deformation[21,22], as well as the ferroelectric phase transition in the presence of various strain and temperature[23]. Unlike the previous studies focusing on demonstration of the super-elasticity in ferroelectric perovskite, we investigated the continuous dynamic process of bended BTO films and the changes in the corresponding ferroelectric patterns and polarization using the phase-field method, which are difficult to observe in experiment.

In this work, we visualized and analyzed the ferroelectric domain evolution via different bending angles of the freestanding BTO film using phase-field simulation. The transformation of domain patterns from the flux-closure to *a*/*c* domains with "vortex-like" structures during *n*-shape (*u*-shape) bending[24] were thoroughly investigated. In addition, we studied the size effect on ferroelectric patterns of bended BTO films and demonstrated that the *a*/*c* phase emerges only as the film becomes thicker than 12 nm, otherwise, *a* phase domain dominates in the freestanding films. Our work provides some insights into the control of the ferroelectric polarization switching process and the dynamic process by adjusting the bending degree of the films.

In the phase-field simulation, the ferroelectric transition and domain structure are described by polarization vector $\boldsymbol{P}=(P_x,P_y,P_z)$. The temporal evolution of the polarization field is solved by the time-dependent Ginzburg–Landau (TDGL) equation:

$$\frac{\partial P_i(\boldsymbol{r},t)}{\partial t} = -L\frac{\delta F}{\delta P_i(\boldsymbol{r},t)}, i = x, y, z, \tag{1}$$

where $\boldsymbol{r}$ is the spatial coordinate, $t$ is the evolution time, $L$ is the kinetic coefficient that is related to the domain evolution, and $F$ is the total free energy that includes the contributions from the Landau energy, the gradient energy, the elastic energy, the electric energy, and flexoelectric coupling energy [25]:

$$F = \iiint_V (f_{Land}(P_i) + f_{grad}(P_{i,j}) + f_{elas}(P_i,\varepsilon_{ij}) + f_{ele}(P_i,E_i) + f_{flexo}(P_i,\varepsilon_{ij},P_{i,j},\varepsilon_{ij,k}))dV. \tag{2}$$

The Landau energy density $f_{Land}$ is given by,

$$f_{Land} = \alpha_{ij}P_iP_j + \alpha_{ijkl}P_iP_jP_kP_l + \alpha_{ijklmn}P_iP_jP_kP_lP_mP_n, \tag{3}$$

where $\alpha_{ij}$, $\alpha_{ijkl}$, $\alpha_{ijklmn}$ are the Landau energy coefficients.

The gradient energy density $f_{grad}$ can be described as follows,

$$f_{grad} = \frac{1}{2}G_{ijkl}P_{i,j}P_{k,l}, \tag{4}$$

where $G_{ijkl}$ are the gradient energy coefficients, and $P_{i,j} = \frac{\partial P_i}{\partial x_j}$.

The elastic energy density $f_{elas}$ can be written as

$$f_{elas} = \frac{1}{2}C_{ijkl}e_{ij}e_{kl} = \frac{1}{2}C_{ijkl}(\varepsilon_{ij} - \varepsilon_{ij}^0)(\varepsilon_{kl} - \varepsilon_{kl}^0), \tag{5}$$

where $C_{ijkl}$ is the elastic stiffness tensor, $e_{ij}$ is the elastic strain, $\varepsilon_{ij}$ and $\varepsilon_{ij}^0$ are the total local strain and eigenstrain, respectively. And $\varepsilon_{ij}^0 = Q_{ijkl}P_kP_l$, where $Q_{ijkl}$ are the electrostrictive coefficients.

The electric energy density $f_{ele}$ is expressed as,

$$f_{ele} = -E_i(P_i + \frac{1}{2}\varepsilon_0\kappa_{ij}E_j), \tag{6}$$

where $E_i$ is the electric field component, $\varepsilon_0$ is the vacuum permittivity, and $\kappa_{ij}$ is the dielectric constant.

And the flexoelectric coupling energy density can be expressed by:

$$f_{flexo} = -\frac{1}{2} f_{ijkl}(\varepsilon_{ij,l}P_k - \varepsilon_{ij}P_{k,l}), \tag{7}$$

where $f_{ijkl}$ is the flexoelectricity tensor. There are three independent flexoelectric coupling coefficients for a material of cubic point group: the longitudinal coupling coefficient $f_{11}$, the transversal coupling coefficient $f_{12}$, and the shear coupling coefficient $f_{44}$. Here, $f_{11}, f_{12}$, and $f_{44}$ are the Voigt form of the flexoelectricity tensor components $f_{1111}, f_{1122}$, and $f_{1212}$, respectively.

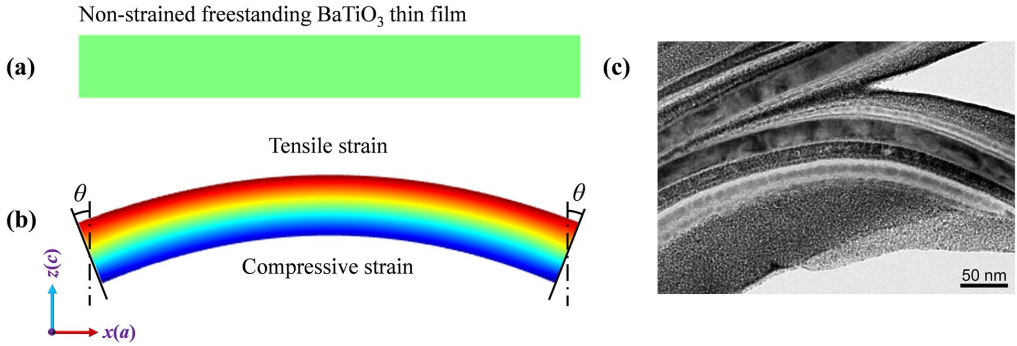

**FIG. 1. Schematic of the freestanding BTO thin film upon bending.** (a) and (b) display the schematics of non-strained and *n*-shape bended BTO thin film, respectively. $\theta$ is defined as bending angle. (c) Cross-sectional TEM image of bended multilayer BTO thin film.

In this work, the related coefficients of BTO in the simulation are taken from the study by Sluka *et al.*[26] (Table S1). The freestanding thin film is discretized at grid size 160 $\Delta x \times 1$ $\Delta x \times 20$ $\Delta x$, where $\Delta x$ is set to 1 nm. The open-circuit condition is applied to the boundaries of the thin film to solve the electrostatic equation, and the temperature is set to be 300 K. When solving the mechanical equilibrium equation, the left and right ends of the film are constrained rigidly as loading grippers. The loading ends are tilted $\theta$ along the neutral surface to achieve the film bending. Then both ends are fixed to stabilize the domain structure evolution. In addition, our simulations are carried out with finite element method. When the BTO film is bended, the top zone and bottom zone are in tensile and compressive strained state, respectively, as shown in Fig. 1(b).

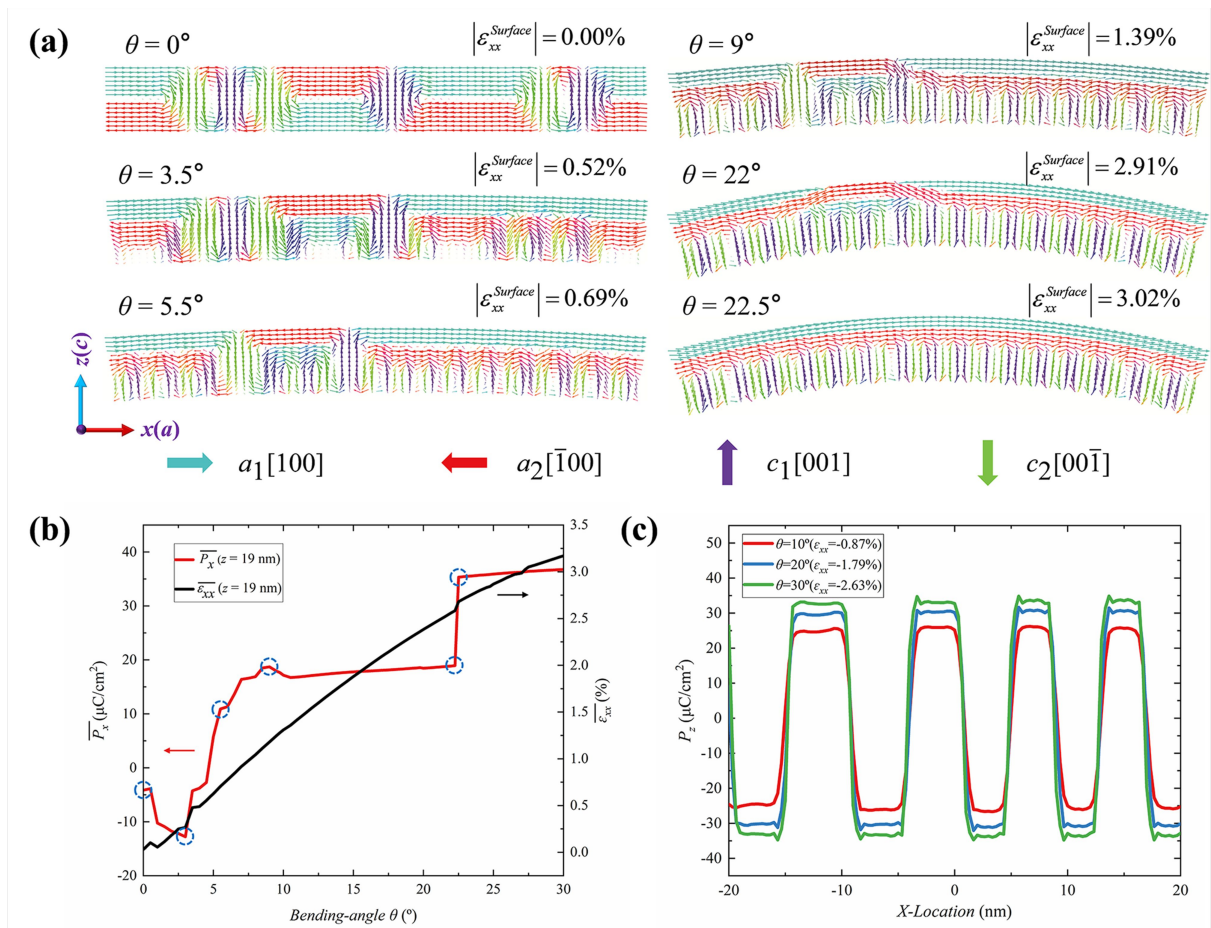

**FIG. 2. Ferroelectric domain structures of the freestanding BTO thin film upon bending.** (a) Domain structures and surface strain of freestanding BTO thin film with a thickness of 20 nm at different bending angles. The cyan, red, purple, and green arrows represent $a_1[100]$, $a_2[\bar{1}00]$, $c_1[001]$ and $c_2[00\bar{1}]$ ferroelectric polarizations, respectively. (b) The dependence of mean polarization component $\overline{P_x}$ and strain $\overline{\varepsilon_{xx}}$ on bending angle $\theta$. Note the $\overline{P_x}$ and $\overline{\varepsilon_{xx}}$ are defined along the line ($z$=19 nm) at the tensile region during $n$-shape bending, the corresponding domain structures at blue-dashed circled states are displayed in (a). (c) The polarization component $P_z$ distribution in the compressed region ($z$=5 nm) of the BTO thin film at three bending angles, 10°, 20°, and 30°.

The bending deformation of ferroelectric thin films significantly influence the transformation of domain structures. When the film is non-strained, the ferroelectric domain structure is the flux-closure pattern to make the surface charge $\boldsymbol{P} \cdot \boldsymbol{n} = 0$ on the free surfaces, which can better minimize the electrostatic energy[27], as shown in Fig. 2(a). As the bending angle increases, the flux-closure domain pattern starts vanishing. Instead, the $a$ domains are generated in the tensile region, due to the stretched in-plane lattice. Similarly, the $c$ domains are generated in the compressive strain region because the in-plane lattices are compressed. Interestingly, "vortex-like" structure occurs consequently at the junction of the $c$ domains and the $a$ domains (near the neutral strain region) due to the rotation of the polarization when the bending angle is about 3.5°. In addition, when $\theta$ changes from 22° to 22.5°, the pattern of 180° multi-domain structure emerges in the tensile-strained region. As the rotation angle is bending from 0º to 22.5º, the corresponding strain of surface changes from 0 to 3.02%. When the stress distribution inside the film is studied, it is found that the stress distribution shows a wave shape at the $a$ and $c$ domain wall region (see Fig. S1). The stress distribution can be influenced by 90º ferroelastic $a/c$ domain wall, however, the 180º non-ferroelastic domain wall will not change the stress distribution during the bending evolution. With the change of domain patterns, the corresponding polarization also changes obviously. Fig. 2(b) demonstrates the variation of the mean polarization component $\overline{P_x}$ and strain $\overline{\varepsilon_{xx}}$ along the line ($z$=19 nm) at the tensile region of BTO thin film with bending angle $\theta$, which is closely related to the transformation of domain patterns during bending.

In the compressive region of BTO thin film, the polarization is aligned either upward or downward, forming a typical multi-domain structure with 180° domain walls. Fig. 2(c) shows the polarization component $P_z$ distribution along the middle line of the compressive region ($z$=5 nm) at three bending angles, 10°, 20°, and 30°. The polarization can increase as the bending angle $\theta$ (the strain of $\varepsilon_{xx}$) increases. It is worth noting that the "spikes" in polarizations are observed at the domain walls. This is due to the influence of the flexoelectric effect, which is responsible for the non-Ising character of a 180° ferroelectric domain wall[28,29].

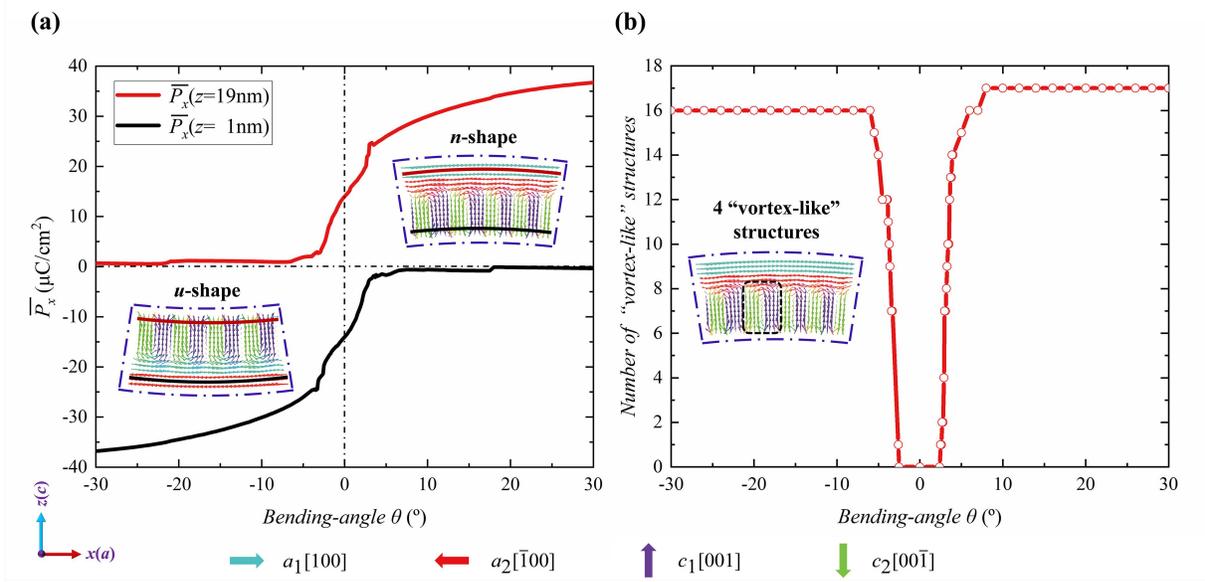

**FIG. 3. Continuous dynamic process of freestanding BTO thin film from *n*-shape bending to *u*-shape bending.** (a) The dependence of mean polarization component $\overline{P_x}$ on bending angle $\theta$. Note that the two lines are close to the top surface ($z$=19 nm) and the bottom surface ($z$=1 nm) of BTO thin film from *n*-shape bending to *u*-shape bending, respectively. (b) Number of "vortex-like" structures with bending angle $\theta$ in the process of changing the bending direction.

To further understand the mechanism of domain structure formation induced by the bending deformation of ferroelectric thin films we investigated the evolution of domain structures from *n*-shape bending to *u*-shape bending. Fig. 3(a) shows the variation of the mean polarization component $\overline{P_x}$ along the lines near the top and bottom interfaces of BTO thin film with bending angle $\theta$ from *n*-shape bending to *u*-shape bending. Here, the bending angle $\theta$ is positive when BTO film under *n*-shape bending. When the film recovers from the *n*-shape bending to the flat state, the mean polarization component $\overline{P_x}$ diminishes gradually. Meanwhile, the domain structure is moving to the left (See detailed analysis in Fig. S3 and Movie S2). When the bending angle $\theta$ reaches 3.5°, the multi-domain structure in the compressive region starts vanishing. Instead, the flux-closure pattern generates (See Fig. S2). Next, when the film is bended downward (*u*-shape bending), the *a* domains are generated in the bottom region. And the *c* domains on the top move to the right as the bending angle gets larger (See Fig. S3 and Movie S2).

With the change of bending angle and bending direction, the number of "vortex-like" structure varies, as shown in Fig. 3(b). The tendency of number change is to decrease first and then increase, which likely be caused by the competition between elastic energy and domain wall energy. For example, the elastic energy increases as the bending angle increases, leading to a decrease on the gradient energy associated with the domain walls to reduce the total energy of the system. Moreover, the chirality of the "vortex-like" structure does not change when the bending direction changes.

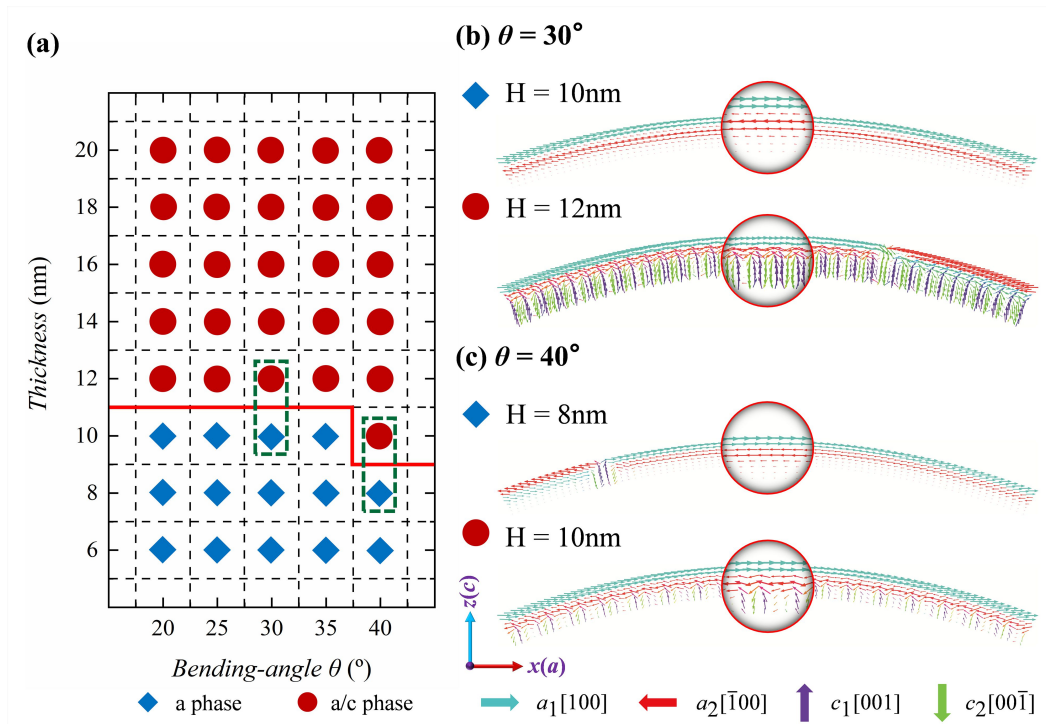

**FIG. 4. The size effect on domain patterns of BTO thin film upon bending.** (a) Domain structures of freestanding BTO thin film with different thicknesses and bending angles. Various domain patterns are denoted by different symbols, blue diamond and red dots indicate $a$ and $a/c$ phase, respectively. (b-c) Domain patterns at critical thicknesses when the bending angle of the BTO thin film are 30° and 40°.

We also analyzed the size effect of the BTO film on domain structures during bending. The diagram of domain patterns with the thickness $H$ and the bending angle $\theta$ of the film was obtained, as shown in Fig. 4(a). When the film is thin, the domain pattern can be regarded as only $a$ phase with multi-domain structure, and the surface polarization is relatively small due to the size effect and the surface effect[30]. When the BTO thin film becomes thicker, the *c* domains appear in the compressive strain zone, indicating the *a/c* phase has formed. For instance, at the bending angle of 30°, the film of 10 nm is *a* phase, while the film of 12nm is *a/c* phase (Fig. 4(b)). The rotation of the polarization in *a/c* phase can largely avoid the mechanical failure[19] by the mismatch stress under bending. It is worth noting that for the BTO film with 10 nm thickness at bending angle of 40°, the domain structure is transformed from *a* phase to *a/c* phase due to sufficient bending strain and large enough thickness.

In summary, we investigated the evolution of ferroelectric polarization in the bended freestanding BTO thin films using phase-field simulations. Domain structures responded continuously to changes in bending-angle (internal strain), and the domain patterns changed in the process of bending. According to our simulation, unlike the domain structures of external-strained BTO films[31], the domain at the compressive region of the bended BTO film showed a contrast difference from that of the tensile region. Therefore, the complex domain structures occur in responsible to the continuous change of the internal strain of the BTO film. Furthermore, we also obtained the phase diagram of domain pattern as a function of the film thickness. The *a/c* phase with "vortex-like" structure emerges only as the film reaching critical thickness, e.g., 12~20 nm. Our simulation showed a light on the distribution of the internal microstructure of bended perovskite oxide films, which could provide guidance for the future study of ferroelectric films on super-elasticity and flexibility.

**Supplementary Information**

See the supplementary material for the stress distribution and domain structures during BTO thin films bending, stress-strain curves, simulation parameters, and BTO thin film bending movies.

**Acknowledgement**

The work at Beijing Institute of Technology is supported by National Key Research and Development Program of China (2019YFA0307900), National Natural Science Foundation of China (Grant Nos. 51972028, 11572040) and Beijing Natural Science Foundation (Grant No. Z190011).

# Supplementary Materials for

# Domain evolution in bended freestanding BaTiO$_3$ ultrathin films: a phase-field simulation

Changqing Guo,[1] Guohua Dong,[2] Ziyao Zhou,[2] Ming Liu,[2] Houbing Huang,[3,4,a)] Jiawang Hong,[1,a)] and Xueyun Wang[1,a)]

[1] School of Aerospace Engineering, Beijing Institute of Technology, Beijing 100081, China

[2] Electronic Materials Research Laboratory, Key Laboratory of the Ministry of Education, School of Electronic and Information Engineering, and State Key Laboratory for Mechanical Behavior of Materials, Xi'an Jiaotong University, Xi'an 710049, China

[3] School of Materials Science and Engineering, Beijing Institute of Technology, Beijing 100081, China

[4] Advanced Research Institute of Multidisciplinary Science, Beijing Institute of Technology, Beijing 100081, China

[a)] Electronic mail: xueyun@bit.edu.cn; hongjw@bit.edu.cn; hbhuang@bit.edu.cn.

**The Supplementary Information files include:**

    Figs. S1-S4
    Table S1
    Movies S1-S2

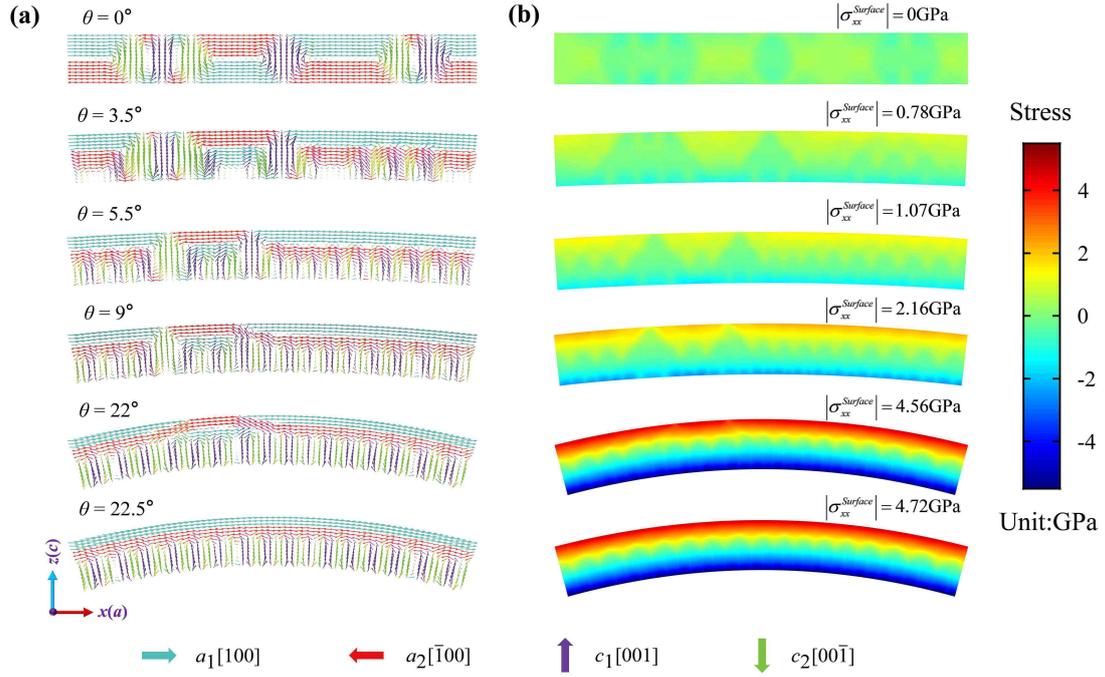

**FIG. S1.** Domain structures and stress distribution of freestanding BTO thin film with a thickness of 20 nm at different bending angles during *n*-shape bending.

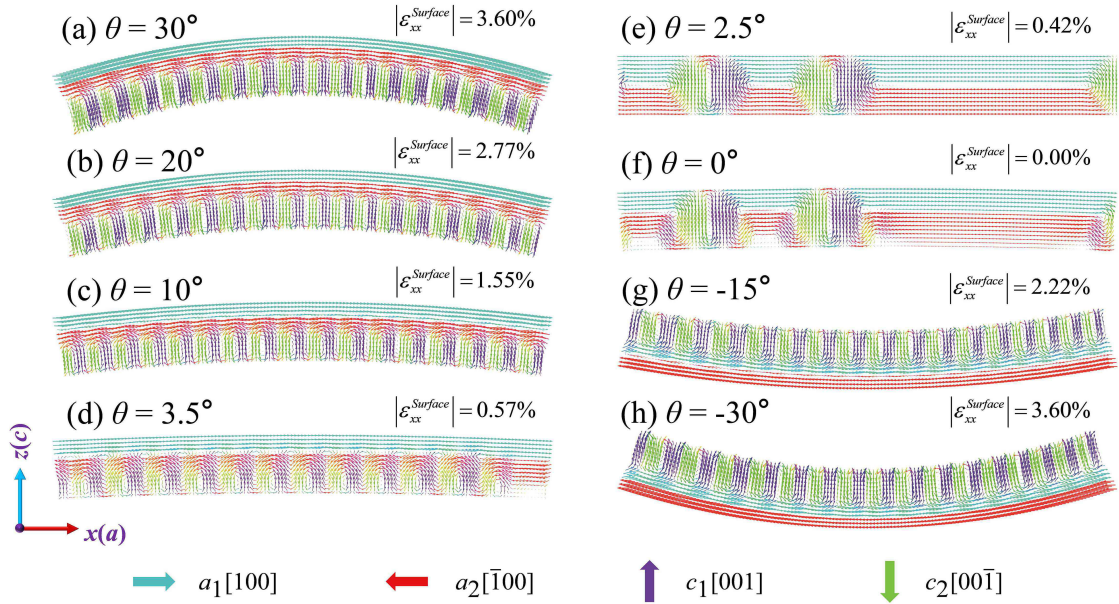

**FIG. S2.** Domain structures of freestanding BTO thin film with a thickness of 20 nm at different bending angles from *n*-shape (a-f) bending to *u*-shape bending (g-h).

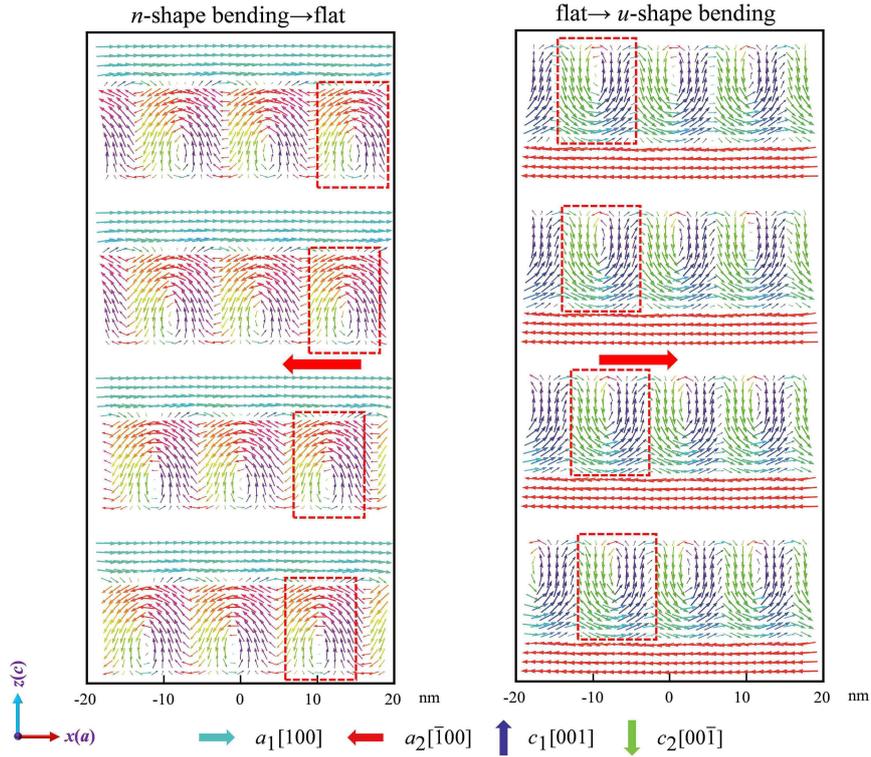

**FIG. S3.** Several local domain structures captured from Movie S2 (Domain structures of freestanding BTO thin film with a thickness of 20 nm from *n*-shape bending to *u*-shape bending).

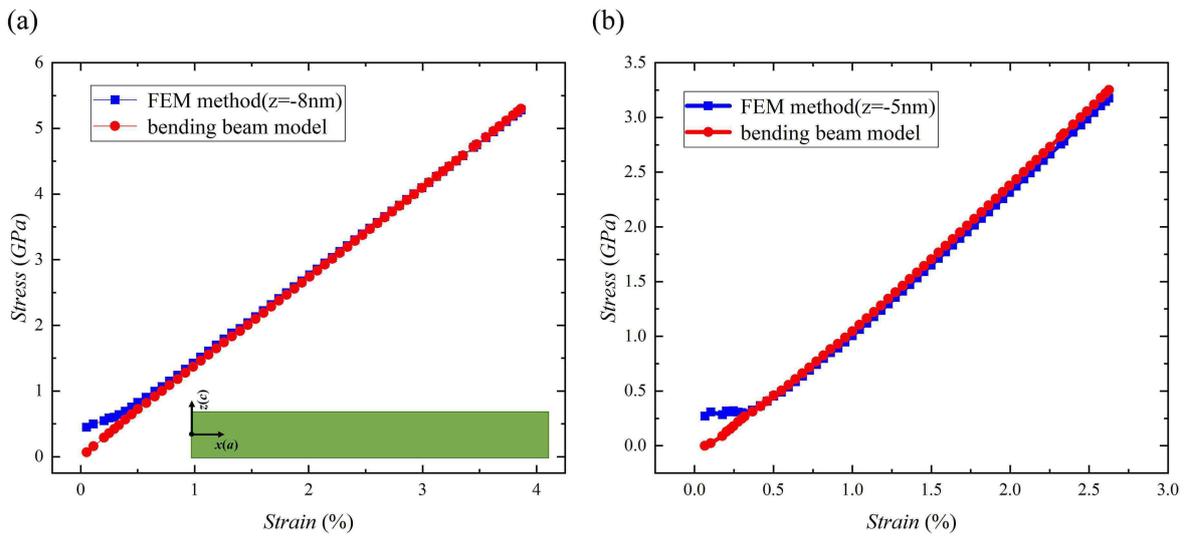

**FIG. S4.** The stress-strain curves at different positions from finite element method and classic bending beam model.

Fig. S3 shows the movement of domain structures of BTO thin film at 20nm from *n*-shape bending to *u*-shape bending. When the bending angle of the film changes from 30° to 0° (from *n*-shape bending to flat state), the domain structure is moving to the left. In contrast, the *c* domains on the top move to the right when BTO film from flat state to *u*-shape bending.

In addition, as shown in Fig. S4, we calculated the stress-strain curves at different positions. The two figures are stress-strain curves with vertical positions of -8 nm and -5 nm, respectively. The blue line represents the stress-strain relationship calculated in our simulations, and the red line is the curve obtained according from the typical bending beam model from classic continuum mechanics. The two lines in the figures overlaps except for the small strain range due to the effect of 90º ferroelastic *a*/*c* domain wall on the stress distribution.

Values of the parameters in the simulation are listed in Table 1. The transversal coupling coefficient $f_{12}$ is set to 2.03V, which is consistent with Dong *et al.*[1] Here, we set $f_{11}=f_{44}=0$ V.

**Table S1.** Values of parameters used in this work[2] (SI units and T in 300 K).

| Parameter | Value | Parameter | Value |
| --- | --- | --- | --- |
| $\alpha_1(J{\cdot}m{\cdot}C^{-2})$ | $-2.7104\times10^7$ | $c_{44}(J{\cdot}m^{-3})$ | $5.43\times10^{10}$ |
| $\alpha_{11}(J{\cdot}m^5{\cdot}C^{-4})$ | $-6.3887\times10^8$ | $Q_{11}(m^4{\cdot}C^{-2})$ | 0.1104 |
| $\alpha_{12}(J{\cdot}m^5{\cdot}C^{-4})$ | $3.23\times10^8$ | $Q_{12}(m^4{\cdot}C^{-2})$ | -0.0452 |
| $\alpha_{111}(J{\cdot}m^9{\cdot}C^{-6})$ | $7.9019\times10^9$ | $Q_{44}(m^4{\cdot}C^{-2})$ | 0.0578 |
| $\alpha_{112}(J{\cdot}m^9{\cdot}C^{-6})$ | $4.47\times10^9$ | $G_{11}(N{\cdot}m^4{\cdot}C^{-2})$ | $5.1\times10^{-10}$ |
| $\alpha_{123}(J{\cdot}m^9{\cdot}C^{-6})$ | $4.91\times10^9$ | $G_{12}(N{\cdot}m^4{\cdot}C^{-2})$ | $-2\times10^{-11}$ |
| $c_{11}(J{\cdot}m^{-3})$ | $2.75\times10^{11}$ | $G_{44}/G'_{44}(N{\cdot}m^4{\cdot}C^{-2})$ | $2\times10^{-11}$ |
| $c_{12}$ (J·m$^{-3}$) | $1.79\times10^{11}$ | $\kappa_b$(1) | 50 |

**Movie S1.** Domain structure evolution under BTO thin film *n*-shape bending.

**Movie S2.** Domain structure evolution when BTO thin film changes bending direction.